\documentclass[11pt, a4paper]{article}
\usepackage{moriond,epsfig}

\def\be{\begin{equation}}
\def\ee{\end{equation}}
\def\bea{\begin{eqnarray}}
\def\eea{\end{eqnarray}}

\begin{document}
\title{Recent Results on non$-D\bar D$ decays of $\psi(3770)$ from BES}
\author{Hai Long Ma (For BES Collaboration)}

\maketitle\abstracts{
BES Collaboration measured the $R$ values at 3.650, 3.6648 and 3.773 GeV,
the $R$ values at 68 energy points in the energy region between 3.650 and
3.872 GeV, the resonance parameters of $\psi(3686)$ and $\psi(3770)$,
the branching fractions for $\psi(3770)\to D^0\bar D^0, D^+D^-, D\bar D$ and
non-$D\bar D$, and
the observed cross sections for some exclusive light hadron final states
at 3.773 and 3.650 GeV. These measurements are made by analyzing the
data sets collected with the BESII detector at the BEPC collider.
}

\vspace{-0.20cm}
\section{Introduction}
\vspace{-0.20cm}
There is a long standing puzzle that the observed cross section
$\sigma_{D\bar D}^{\rm obs}$ for $D\bar D$ production at the
$\psi(3770)$ peak is less than the observed cross section
$\sigma_{\psi(3770)}^{\rm obs}$ for $\psi(3770)$ production \cite{rzhc}.
Precise or direct measurements of the $R$ values at 3.650, 3.6648 and
3.773 GeV \cite{3r}, the $R$ values in the energy region between 3.650
and 3.872 GeV \cite{68r}, the resonance parameters of $\psi(3686)$ and
$\psi(3770)$ \cite{brdd,resr}, the branching fractions for the $\psi(3770)$
decays \cite{3r,brdd} and the observed cross sections for more exclusive
light hadron final states produced in $e^+e^-$ annihilation at 3.773 and
3.650 GeV \cite{crs_light_hads} are important to understand the discrepancy.
In addition,
precise measurements of the $R$ values are also important to test the
validity of the pQCD calculation in this energy region and to calculate the
effects of vacuum polarization on the parameters of the standard model
\cite{3r,68r}. In this paper, we report the results of these measurements.

For convenience, we call the data taken at the c.m. (center-of-mass) energies of
3.650, 3.6648 and 3.773 GeV, the data taken at 49 energy points in the energy
region between 3.660 and 3.872 GeV in March 2003,
the data taken at 68 energy points in the energy region between
3.650 and 3.872 GeV in December 2003 to be the data A, the data B and the
data C, respectively.

\vspace{-0.20cm}
\section{Measurements of $R$ values}
\vspace{-0.20cm}
With the data A, we measured the lowest order cross sections $\sigma_{\rm
had}^{\rm obs}$ and the $R$ values ($R=
\sigma^0_{e^+e^-\to {\rm hadrons}}/\sigma^0_{e^+e^-\to \mu^+\mu^-}$) for
inclusive hadron production at 3.650, 3.6648 and 3.773 GeV. These results \cite{3r} are
summarized in Table \ref{tab:r_val}, where the first error is the combined
statistical and point-to-point systematic error, and the second is the
common systematic.

\begin{table}[htbp]
\begin{center}
\caption{The measured $\sigma_{\rm had}^{\rm obs}$ and $R$ values for
inclusive hadron production at 3.650, 3.6648 and 3.773 GeV.
\label{tab:r_val}}
\begin{tabular}{|c|c|c|c|} \hline
$E_{\rm cm}$ (GeV)              &3.6500               &3.6648               &3.7730               \\ \hline
$\sigma_{\rm had}^{\rm obs}$[nb]&$18.98\pm0.20\pm0.76$&$18.30\pm0.27\pm0.73$&$27.68\pm0.27\pm1.38$\\
$R_{\rm had}$                   &$2.26\pm0.02\pm0.09$ &$2.31\pm0.03\pm0.09$ &$3.75\pm0.04\pm0.19$ \\
$R_{\rm uds}$                   &$2.24\pm0.02\pm0.09$ &$2.19\pm0.03\pm0.09$ & -                   \\
$R_{\rm uds+\psi(3770)}$        &-                    &-                    &$3.75\pm0.04\pm0.19$ \\
\hline
\end{tabular}
\end{center}
\end{table}

\begin{figure}[htbp]
\begin{center}
\psfig{figure=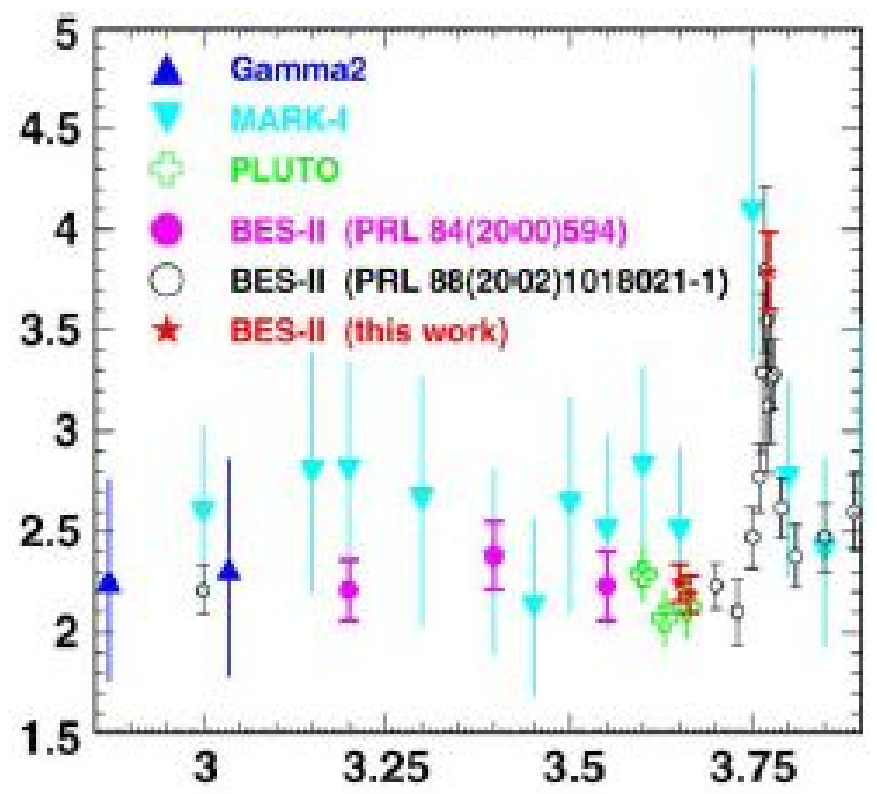,width=4cm,height=4.0cm}
\put(-80,-10.0){\bf $E_{\rm cm}$ (GeV)}
\put(-135,-10.0){\rotatebox{90}{\bf $R_{\rm uds(c)+\psi(3770)}(s)$ or $R_{\rm had}(s)$}}
\put(-20,100){\bf (a)}
\hspace{2.5cm}
\psfig{figure=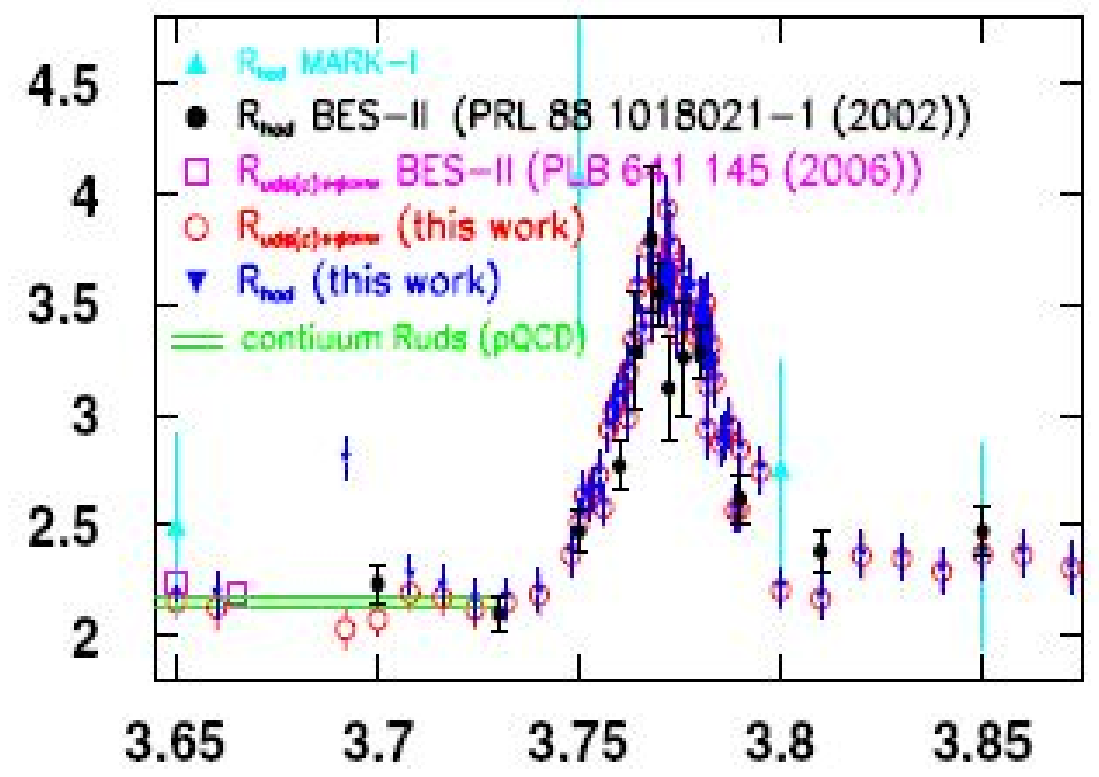,width=4cm,height=4.0cm}
\put(-80,-10.0){\bf $E_{\rm cm}$ (GeV)}
\put(-135,-10.0){\rotatebox{90}{\bf $R_{\rm uds(c)+\psi(3770)}(s)$ or $R_{\rm had}(s)$}}
\put(-20,100){\bf (b)}
\caption{The measured $R$ values (a) at 3.650, 3.6648 and 3.773 GeV
and (b) at 68 energy points in the energy region between 3.650 and 3.872 GeV,
compared with the other measurements.
\label{fig:r_value}}
\end{center}
\end{figure}

With the data C, we also measured the continuum $R_{\rm uds}$ below the
$D\bar D$ production, the $R_{\rm uds(c)+\psi(3770)}(s)$ and the
$R_{\rm had}(s)$ values in $e^+e^-$ annihilation at all of the 68 energy points \cite{68r}. They are compared with the other measurements
in Fig. \ref{fig:r_value}.

\vspace{-0.20cm}
\section{Measurements of resonance parameters of $\psi(3686)$ and
$\psi(3770)$}
\begin{figure}[htbp]
\begin{center}
\psfig{figure=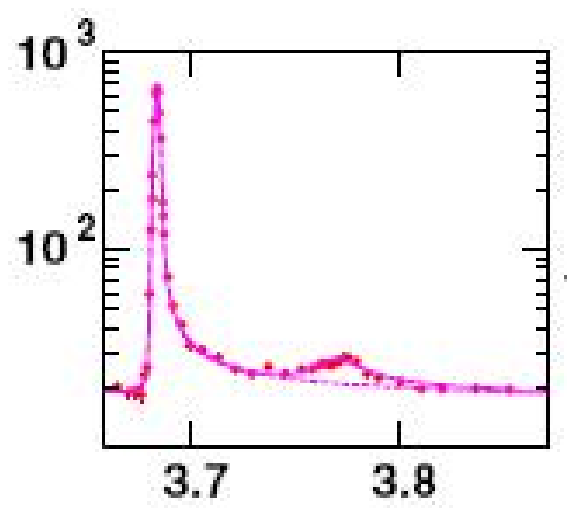,width=4cm,height=4.0cm}
\put(-80,-10.0){\bf $E_{\rm cm}$ (GeV)}
\put(-135,40.0){\rotatebox{90}{\bf $\sigma_{\rm had}^{\rm obs}$ (nb)}}
\put(-25,95){\bf (a)}
\hspace{1.0cm}
\psfig{figure=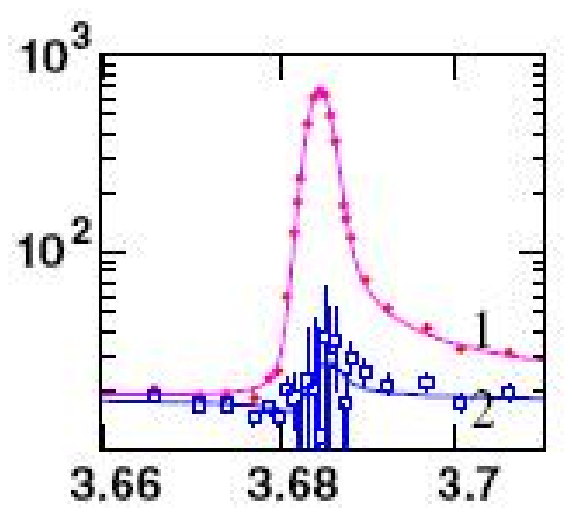,width=4cm,height=4.0cm}
\put(-80,-10.0){\bf $E_{\rm cm}$ (GeV)}
\put(-135,40.0){\rotatebox{90}{\bf $\sigma_{\rm had}^{\rm obs}$ (nb)}}
\put(-25,95){\bf (b)}
\hspace{1.0cm}
\psfig{figure=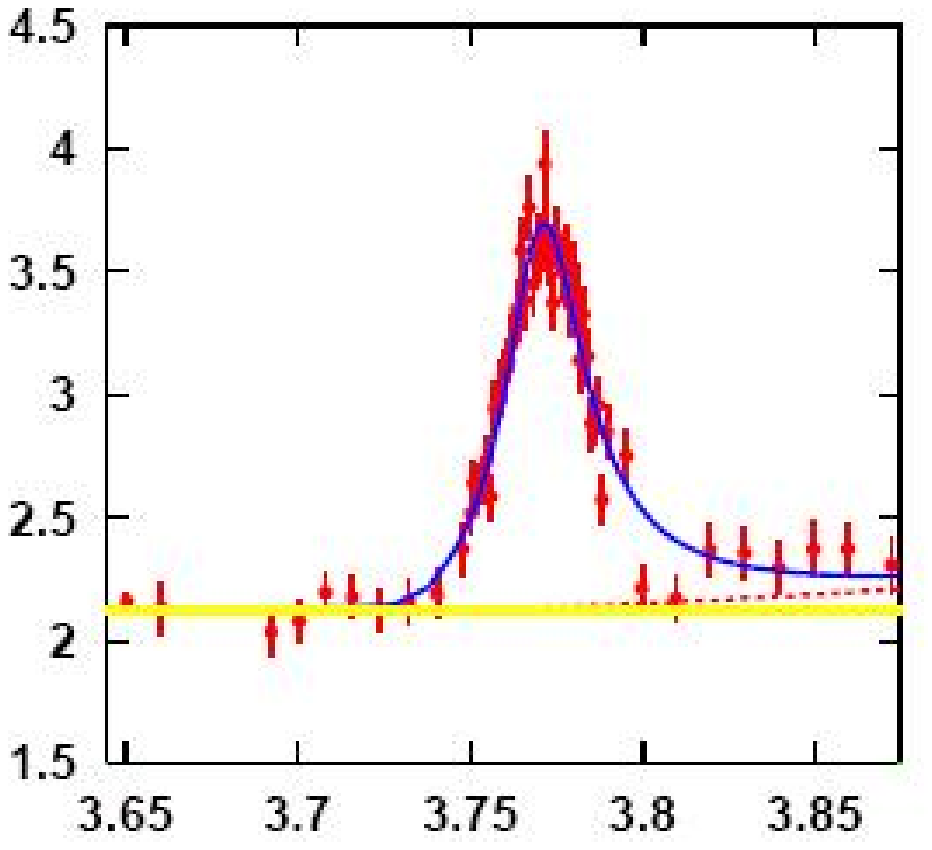,width=4cm,height=4.0cm}
\put(-80,-10.0){\bf $E_{\rm cm}$ (GeV)}
\put(-135,20.0){\rotatebox{90}{\bf $R_{\rm uds(c)+\psi(3770)}(s)$}}
\put(-20,100){\bf (c)}
\caption{The fits to the observed cross sections $\sigma_{\rm had}^{\rm obs}$
and the $R_{\rm uds(c)+\psi(3770)}$ values versus $E_{\rm cm}$, where the
lines show the fits, (a) and (b) are from the data B and (c) is from the
data C.
\label{fig:fitcrs}}
\end{center}
\end{figure}

\vspace{-0.20cm}
\label{resonance}
A better way to measure the branching fractions for $\psi(3770)\to D\bar D$
is to simultaneously analyze the energy-dependent cross sections for the
inclusive hadron, $D^0\bar D^0$ and $D^+D^-$ production in the energy range
covering both the $\psi(3686)$ and $\psi(3770)$. We first
accurately measured the resonance parameters of the two
resonances \cite{brdd} by fitting the observed cross sections at 49 energy
points from the data B. In addition, we reported precision measurements of
the mass, the total width and the partial leptonic width of the
$\psi(3770)$ by further analyzing the measured $R$ values at 68 energy points
\cite{resr} from the data C.
The fits to the energy-dependent observed cross sections $\sigma_{\rm
had}^{\rm obs}$ and the $R$ values for inclusive hadron production are shown in Fig.
\ref{fig:fitcrs}. The fitted results are summarized in Table
\ref{tab:fit}.

\begin{table}[htbp]
\begin{center}
\caption{Summary of the measured resonance parameters of the $\psi(3770)$
and $\psi(3686)$, where M is the mass, $\Gamma^{\rm tot}$ is the total
width, $\Gamma^{ee}$ is the partial leptonic width and $\Delta \rm M$ is the
mass difference of the $\psi(3770)$ and $\psi(3686)$.
\label{tab:fit}}
\begin{tabular}{|c|c|c|c|c|c|} \hline
Res.        &Ref.& M(MeV)             &$\Gamma^{\rm tot}$(MeV)&$\Gamma^{ee}$(eV)&$\Delta\rm M$(MeV)\\ \hline
$\psi(3770)^{\rm B}$&\cite{brdd}&$3772.2\pm0.7\pm0.3$&$26.9\pm2.4\pm0.3$     &$251\pm26\pm11$  &                  \\
$\psi(3686)^{\rm B}$&\cite{brdd}&$3685.5\pm0.0\pm0.3$&$0.331\pm0.058\pm0.002$&$2330\pm36\pm110$&$86.7\pm0.7$      \\ \hline
$\psi(3770)^{\rm C}$&\cite{resr}&$3772.4\pm0.4\pm0.3$&$28.6\pm1.2\pm0.2$     &$279\pm11\pm13$  &                  \\ \hline
\end{tabular}
\end{center}
\end{table}

\vspace{-0.20cm}
\section{Determinations of $R_{\rm uds}$ and $\sigma_{\psi(3770)}$}
\vspace{-0.20cm}
Averaging the measured $R_{\rm uds}$ values at 3.650 and 3.6648 GeV from the
data A listed
in Table \ref{tab:r_val} by the combined statistical and point-to-point
systematic error, we obtain $\bar R_{\rm uds}=2.218\pm0.019\pm0.089,$
where the first error is the combined statistical and point-to-point
systematic error, and the second is the common systematic.

Fitting to the energy-dependent $\sigma_{\rm had}^{\rm obs}$ from the data B yields
$R_{\rm uds}=2.262\pm0.054\pm0.109$ in the energy region between 3.660
and 3.872 GeV. Fitting to the measured $R$ values at 68 energy points
from the data C yields $R_{\rm uds}=2.121\pm0.023\pm0.084$ in the energy region
between 3.650 and 3.872 GeV. Here, the errors are statistical and
systematic, respectively.

Ignoring the contribution from the continuum $D\bar D$
production at the $\psi(3770)$ peak, we obtained from the data A to be
$R_{\psi(3770)}=1.528\pm0.042\pm0.131$ due to $\psi(3770)$ decay into
hadrons with the measured $R_{\rm uds+\psi(3770)}$ value at 3.773 GeV
as listed in Table \ref{tab:r_val} and $\bar R_{\rm uds}$ measured below
$D\bar D$ threshold. This lead to the lowest order cross section
$\sigma_{\psi(3770)}^{\rm 0}$ and the observed cross section
$\sigma_{\psi(3770)}^{\rm obs}$ for $\psi(3770)$ production at
3.773 GeV. The resonance parameters of the $\psi(3770)$ obtained by
fitting to the energy-dependent $\sigma_{\rm had}^{\rm obs}$
or fitting to the $R_{\rm uds(c)+\psi(3770)}$ values from analyzing the cross section scan data samples
can further give $\sigma_{\psi(3770)}^{\rm 0}$ and $\sigma_{\psi(3770)}
^{\rm obs}$ at $\psi(3770)$ peak. They are summarized in Table
\ref{tab:psipp}.

\begin{table}[htbp]
\begin{center}
\caption{Summary of the measured $\sigma_{\psi(3770)}^{\rm 0}$
and $\sigma_{\psi(3770)}^{\rm obs}$ at 3.773 GeV or its peak.
\label{tab:psipp}}
\begin{tabular}{|c|c|c|c|} \hline
Data Sample&Ref.&$\sigma_{\psi(3770)}^{\rm 0}$[nb]&$\sigma_{\psi(3770)}^{\rm obs}$[nb]\\ \hline
A&\cite{3r}  &$9.323\pm0.103\pm0.801$          &$7.179\pm0.195\pm0.630$            \\
B&\cite{brdd}&$9.63\pm0.66\pm0.35$             &$6.94\pm0.48\pm0.28$               \\
C&\cite{resr}&$10.06\pm0.37\pm0.43$            &$7.25\pm0.27\pm0.34$               \\ \hline
\end{tabular}
\end{center}
\label{psipp}
\end{table}

\vspace{-0.20cm}
\section{Measurements of branching fractions for $\psi(3770)\to
\vspace{-0.20cm}
D^0\bar D^0$, $D^+D^-$, $D\bar D$ and non$-D\bar D$}
Assuming that there are no other new structures and effects except the
$\psi(3770)$ resonance and the continuum hadron production in the energy
region from 3.70 to 3.86 GeV, we can determine \cite{3r} the branching
fractions for $\psi(3770)\to D^0\bar D^0$, $D^+D^-$, $D\bar D$ and
non$-D\bar D$ with the measured $\sigma_{\psi(3770)}^{\rm obs}$ and
the observed cross sections \cite{dcrs1,dcrs2} $\sigma^{\rm obs}_{D^0
\bar D^0}$, $\sigma^{\rm obs}_{D^+ D^-}$ and $\sigma^{\rm obs}_{D\bar
D}$ for $D^0\bar D^0$,
$D^+D^-$ and $D\bar D$ production measured by analyzing the same data
sample A. Fitting to the energy-dependent $\sigma_{\rm had}^{\rm obs}$,
$\sigma^{\rm obs}_{D^0\bar D^0}$, $\sigma^{\rm obs}_{D^+ D^-}$ and
$\sigma^{\rm obs}_{D\bar D}$ in the energy range
covering both the $\psi(3686)$ and $\psi(3770)$ from the data B, we can also measured
the branching fractions for $\psi(3770)\to D^0\bar D^0$, $D^+D^-$, $D\bar
D$ and non$-D\bar D$. The measured branching fractions
are summarized in Table \ref{tab:brdd}.

\begin{figure}[htbp]
\begin{center}
\psfig{figure=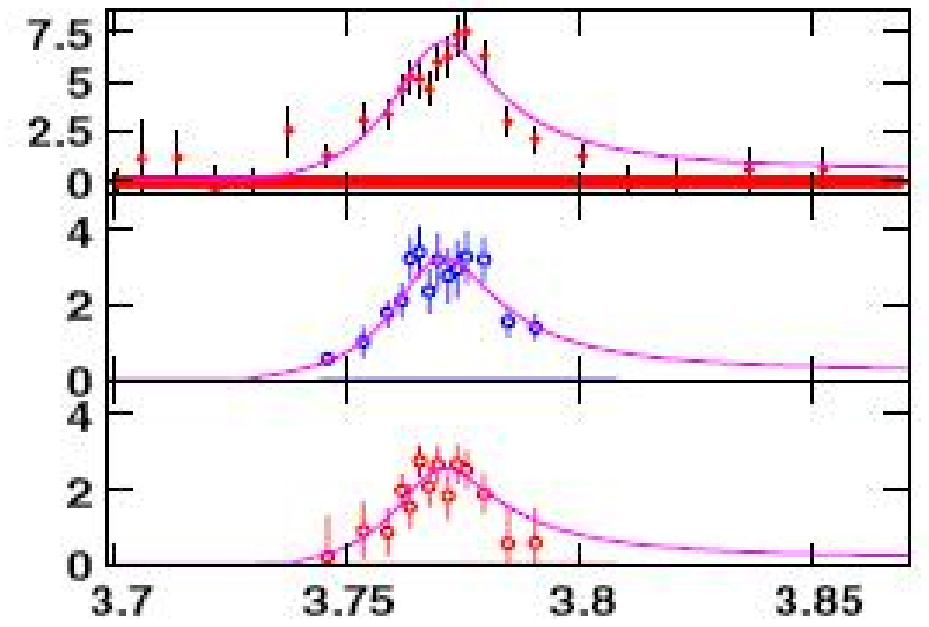,width=4cm,height=4.0cm}
\put(-80,-10.0){\bf $E_{\rm cm}$ (GeV)}
\put(-135,40.0){\rotatebox{90}{\bf $\sigma^{\rm obs}$ (nb)}}
\put(-20,100){\bf (a)}
\put(-20,65){\bf (b)}
\put(-20,30){\bf (c)}
\caption{The observed cross sections versus $E_{\rm cm}$ with fits, where
(a) shows inclusive hadron cross section, (b) and (c) show the 
$D^0\bar D^0$ and $D^+D^-$ cross sections, respectively.
\label{fig:ddbar}}
\end{center}
\end{figure}

\begin{table}[htbp]
\begin{center}
\caption{The measured branching fractions for $\psi(3770)\to D^0\bar D^0$,
$D^+D^-$, $D\bar D$ and non-$D\bar D$.
\label{tab:brdd}}
\begin{tabular}{|c|c|c|c|c|c|} \hline
$\psi(3770)\to$   &Ref.       &$D^0\bar D^0$     &$D^+D^-$          &$D\bar D$        &non-$D\bar D$      \\ \hline
$\mathcal B^{\rm A}$[\%]&\cite{3r}  &$49.9\pm1.3\pm3.8$&$35.7\pm1.1\pm3.4$&$85.5\pm1.7\pm5.8$&$14.5\pm1.7\pm5.8$\\
$\mathcal B^{\rm B}$[\%]&\cite{brdd}&$46.7\pm4.7\pm2.3$&$36.9\pm3.7\pm2.8$&$83.6\pm7.3\pm4.2$&$16.4\pm7.3\pm4.2$\\ \hline
\end{tabular}
\end{center}
\end{table}

\vspace{-0.20cm}
\section{Measurements of the observed cross sections for some exclusive light
hadron final states produced in $e^+e^-$ annihilation at 3.773 and 3.650 GeV}
\vspace{-0.20cm}
We measured the observed cross sections for
the exclusive light hadron final states of $\phi\pi^0$, $\phi\eta$,
$\phi\pi^+\pi^-$, $\phi K^+K^-$,
$\phi p\bar p$, $2(\pi^+\pi^-)\eta$, $2(\pi^+\pi^-)\pi^0$,
$K^+K^-\pi^+\pi^-\pi^0$, $2(K^+K^-)\pi^0$, $p \bar p \pi^0$,
$p \bar p\pi^+\pi^-\pi^0$ and $3(\pi^+\pi^-)\pi^0$ produced in $e^+e^-$
annihilation at $\sqrt{s}=$ 3.773 and 3.650 GeV. The preliminary results
\cite{crs_light_hads} are
shown in Table \ref{tab:crs_light_hadrons}, where the upper
limits are set at 90\% C.L..
We ignore the interference effects between the continuum and
resonance amplitudes, since we do not know the details about the
two amplitudes. Therefore we can not draw a conclusion that the
$\psi(3770)$ does not decay into these final states even if we do not
observe significant difference between the observed cross sections
for most light hadron final states at the two energy points.

{\small
\begin{table*}[htbp]
\begin{center}
\caption{Comparisons of the observed cross sections for $e^+e^-\to$
exclusive light hadrons at 3.773 and 3.650 GeV.
\label{tab:crs_light_hadrons}}
\begin{tabular}{|l|c|c|} \hline
Final State   & $\sigma^{(\rm up)}_{e^+e^-\to f}$(@3.773 GeV)[pb]
             & $\sigma^{(\rm up)}_{e^+e^-\to f}$(@3.650 GeV)[pb]    \\ \hline
$\phi\pi^0$             &$<3.5$               &$<8.9$               \\
$\phi\eta$              &$<12.6$              &$<18.0$              \\
$\phi\pi^+\pi^-$        &$<11.1$              &$<22.9$              \\
$\phi K^+K^-$           &$15.8 \pm5.1 \pm1.8$ &$17.4 \pm9.2 \pm 2.0$\\
$\phi p \bar p$         &$<5.8$               &$<9.1$               \\
$2(\pi^+\pi^-)\eta$     &$153.7\pm40.1\pm18.4$&$86.6 \pm40.3\pm10.4$\\
$2(\pi^+\pi^-)\pi^0$    &$80.9 \pm13.9\pm10.0$&$124.3\pm21.7\pm14.9$\\
$K^+K^-\pi^+\pi^-\pi^0$ &$171.6\pm26.0\pm20.9$&$222.8\pm37.7\pm27.2$\\
$2(K^+K^-)\pi^0$        &$18.1 \pm7.7 \pm2.1$ &$<23.0$              \\
$p\bar p\pi^0$          &$10.1 \pm2.2 \pm1.0$ &$9.2  \pm3.4 \pm 1.0$\\
$p\bar p\pi^+\pi^-\pi^0$&$53.1 \pm9.2 \pm6.8$ &$29.0 \pm11.1\pm 3.7$\\
$3(\pi^+\pi^-)\pi^0$    &$105.8\pm34.4\pm16.9$&$126.6\pm47.1\pm19.2$\\ \hline
\end{tabular}
\end{center}
\end{table*}
}

\vspace{-0.20cm}
\section{Summary}
\vspace{-0.20cm}
Using the data sets collected with the BESII detector at
the BEPC collider, BES Collaboration measured
the $R$ values at 3.650, 3.6648 and 3.773 GeV,
the $R$ values in the energy region between 3.650 and 3.872 GeV,
the resonance parameters of $\psi(3686)$ and $\psi(3770)$,
the branching fractions for $\psi(3770)\to D^0\bar D^0, D^+D^-, D\bar D$ and
non-$D\bar D$, and
the observed cross sections for some exclusive light hadron final states
at 3.773 and 3.650 GeV.

\end{document}